\documentclass[]{emulateapj}
\usepackage{natbib}
\usepackage{epsfig}
\usepackage{graphics,graphicx}
\usepackage{rotating}
\usepackage{lscape}

\newcommand{\csn}{$\chi^2_\nu$}
\newcommand{\egcms}{ergs cm$^{-2}$ s$^{-1}$}
\newcommand{\cms}{cm$^{-3}$ s}

\newcommand{\emax}{$E_{\rm max}$}
\newcommand{\mg}{$\mu G$}

\newcommand{\pshock}{{PSHOCK}}

\newcommand{\vpshock}{{VPSHOCK}}

\newcommand{\srcut}{{SRCUT}}
\newcommand{\sresc}{{SRESC}}

\newcommand{\limb}{{\it limb}}
\newcommand{\cp}{{\it cap}}
\newcommand{\ctr}{{\it center}}

\newcommand{\asca}{{\sl ASCA}}
\newcommand{\xmm}{{\sl XMM-Newton}}

\newcommand{\rosat}{{\sl ROSAT}}
\newcommand{\rxte}{{\sl RXTE}}
\newcommand{\beppo}{{\sl BeppoSAX}}
\newcommand{\chandra}{{\sl Chandra}}

\newcommand{\sis}{SIS}
\newcommand{\gis}{GIS}

\newcommand{\pca}{PCA}

\newcommand{\rolloff}{$\nu_{\rm rolloff}$}
\newcommand{\norm}{{\it norm}}

\shortauthors{Dyer et al.}
\shorttitle{Thermal and Non-Thermal X-Rays II:SN~1006}
                                                       
\received{2003 July 25}

\begin{document}

\title{Separating Thermal and Non-Thermal X-Rays in Supernova Remnants
II: Spatially Resolved Fits to SN~1006 AD}

\author{K.K. Dyer\altaffilmark{1}} \affil{National Radio Astronomy Observatory,
P.O. Box O, Socorro NM 87801}

\altaffiltext{1}{NSF Astronomy and Astrophysics Postdoctoral Fellow}

\author{S.P. Reynolds, K.J. Borkowski} \affil{North Carolina State
University, Physics Dept. Box 8202, Raleigh NC 27695-8202}

\email{kdyer@nrao.edu}

\begin{abstract}

We present a spatially resolved spectral analysis of full \asca\/
observations of the remnant of the supernova of 1006 AD.  This remnant
shows both nonthermal X-ray emission from bright limbs, generally
interpreted as synchrotron emission from the loss-steepened tail of
the nonthermal electron population also responsible for radio
emission, and thermal emission from elsewhere in the remnant.
In
earlier work, we showed that the spatially integrated spectrum was
well described by a theoretical synchrotron model in which shock
acceleration of electrons was limited by escape, in combination with
thermal models indicating high levels of iron from ejecta.  Here we
use new spatially resolved subsets of the earlier theoretical
nonthermal models for the analysis.  We find that emission from the
bright limbs remains well described by those models, and refine the
values for the characteristic break frequency.  We show that
differences between the northeast and southwest nonthermal limbs are small, too
small to account easily for the presence of the northeast limb, but not the
southwest, in TeV gamma-rays.  Comparison of spectra of the nonthermal limbs
and other regions confirms that simple cylindrically symmetric
nonthermal models cannot describe the emission, and we put limits on
nonthermal contributions to emission from the center and the northwest and southeast
limbs. We can rule out solar-abundance models in all regions, finding evidence
for elevated abundances. However, more sophisticated models will be
required to accurately characterize these abundances.


\end{abstract}

\section{Introduction}


A rapidly growing number of galactic remnants exhibit nonthermal (and
non-pulsar-related) emission in the X-ray band. Nonthermal emission
can dominate over much or all of the remnant, as is the case with
SN~1006 \citep{Dyer2000}, G266.2--1.2 [RX J0852.0--4622]
\citep{Slane6}, G347.3--0.5 \citep{Slane1999, Uchiyama2003}, and AX J1843.8-0352, \citep{Ueno2003}. 
Alternatively an SNR can have signatures of nonthermal emission along with
X-ray spectral lines, for example 
 RCW 86 \citep{Rho2003}, Cassiopeia A
\citep{Vink2003a}, and Tycho, \citep{Hwang2002}. 
Finally some SNRs
(Cassiopeia A, Kepler, Tycho, SN~1006 and RCW~86) reveal nonthermal
emission in the hard-X-ray
sensitivity of the \rxte\/ \pca\/ (up to 60 keV), suggesting
nonthermal emission may be common among most young SNRs
\citep{Allen1999}.

 In most of these instances, a strong case can be made that the
 nonthermal emission is synchrotron radiation.  The total number of
 SNRs with suspected nonthermal emission is now in the double digits
 and we expect, as new X-ray instruments improve in spectral and
 spatial resolution, many more SNRs will be found with varying amounts
 of X-ray synchrotron emission, previously undetected among thermal
 continuum and line emission. In the past this emission was missed
 because few analyses included models for nonthermal emission. Even of
 those that do, many do not take advantage of this unique opportunity
 to extract information about particle acceleration from the emission
 of ultra-relativistic electrons. We know that the spectrum
 extrapolated from radio frequencies must roll off before X-ray
 energies to avoid exceeding X-ray flux measurements (Reynolds \&
 Keohane 1999; Hendrick
\& Reynolds 2001).  Therefore observations of X-ray synchrotron 
emission lie in
the regime where the particle spectrum is changing
rapidly. Shock-acceleration models that can reproduce this drop-off
may be able to provide information about the remnant age, radiative
losses of electrons or the spectrum of magnetohydrodynamic (MHD) waves
near the shock.


In Dyer et al.~(2000; hereafter Paper I) we demonstrated that the
total-flux spectrum from SN~1006 was well fit by a combination of new
synchrotron models and plane-shock thermal models, and that the fit was an
improvement over previous fits both due to the robustness of the
synchrotron models (which use radio measurements to constrain the
nonthermal emission) and due to the new accuracy of the thermal models,
which, with the assistance of the synchrotron model, detected for the
first time half a solar mass of iron. However, these models were fit to
spectra summed over the entire remnant and it is well known that the
remnant spectra can vary significantly across the face of a remnant.  In
particular \citet{Koyama6} settled a controversy about the X-ray spectrum of
SN~1006 by exhibiting a {\em difference} in spectra between the limb and
center.  

One of the most compelling reasons for spatially resolved studies of
SN~1006 is the nonaxisymmetric TeV $\gamma$-ray emission from SN~1006
\citep{Tanimori1998}. In Paper I we interpreted this emission as 
inverse-Compton upscattering of cosmic-microwave-background photons
and derived a preshock magnetic field of $\sim$3 \mg,
implying a mean field in the remnant of about 9 \mg.
Since then, full
analysis of the $\gamma$-ray spectrum yielded a magnetic field of 4~\mg\/ \citep{Tanimori2001}, in reasonable agreement with our previous
results. Inverse-Compton is not the only possible emission mechanism to
explain the $\gamma$-rays -- recent work
\citep{Berezhko2002} attributes the TeV emission from SN~1006 to decay of $\pi^0$
mesons produced by inelastic collisions between cosmic-ray protons and
thermal gas. However, neither interpretation explains why $\gamma$-rays
are seen only in the northeast (NE) -- only spatially resolved studies
could hope to answer this question.

The clear symmetry of SN~1006 about a northwest-southeast axis suggests that
generalizations of the cylindrically symmetric model used in Paper I
(\sresc\/) is a reasonable starting point for such studies.  
While the symmetric synchrotron model 
is an oversimplification,
the results from the whole remnant fits (Paper I) were very encouraging,
suggesting that we extend the model to find its limits.  We hope to
answer a set of related questions:  Is the spatial distribution of
emission from SN~1006 consistent with an axially symmetric model?
How exact is the bilateral symmetry, and can any hint be found in
the X-rays for the gross asymmetry shown by the TeV $\gamma$-ray
emission?  In the regions where nonthermal emission dominates can
thermal emission be characterized by simple models?

There has been much recent work on SN~1006. \citet{Vink2000} analyzed
integrated spectra from \beppo\/ of SN~1006 with a new version of SPEX
designed to simultaneously fit regions which overlap in the point
spread function. They found an adequate two temperature fit to the
thermal emission and supersolar abundances.  New high resolution
\chandra\/ observations of the NW and NE provide the best spatial resolution images to
date
and several groups have been exploring the fine
details. \citet{Long2003} analyzed the first two
\chandra\/ pointings and found no evidence for the halo predicted
by \citet{Reynolds1996}. The radio and X-ray features were found to be
perfectly correlated in the NE although above 0.8 keV X-ray limb
brightening is more pronounced. Clumps of thermal X-ray material were
found interior to the shock in the NW and NE with super-solar
abundances in the NW. \citet{Bamba2003b} carried out
spatial-spectral fits of cross sections of the \chandra\/ observation
of the sharp fine filaments in the NE. Using an energy cut to separate
thermal and nonthermal emission they found the nonthermal emission to
be narrower than the thermal emission. From thermal fits they found
super-solar abundances, including iron, and a profile consistent with
Sedov dynamics. More recent
\chandra\/ observations by Hughes use short exposures to image 
the entire SNR.

Here we attempt to extract the maximum useful information from all
\asca\/ observations of SN~1006.  We present spatial subsets of the
\sresc\/ synchrotron model of Paper I, and apply them
to describe the nonthermal emission.  We believe that this study
represents the fullest use of the substantial amount of observing time
obtained with the \asca\/ satellite on SN~1006, and that further advances
will require both better data and better thermal and nonthermal models.

\section{The context of this work}
\label{context}

Much ground has been covered in the process of moving from
phenomenological power laws to the development of a synchrotron model
appropriate for regions of SN~1006.  \citet{Reynolds1999} used a
maximally curved model, \srcut \  to find upper limits for synchrotron
emission in Galactic remnants. The model was fit, ignoring evidence
for thermal emission, as if all X-ray emission were synchrotron. Since
the model had maximal curvature, this procedure produced the
highest electron energy at which the lower-energy power-law could
begin to steepen and still not exceed observed X-rays, 
placing a solid upper limit on
the energy to which SNRs could accelerate electrons with the same
slope as at radio emitting energies. The results were significant --
even if all X-ray emission were synchrotron these Galactic SNRs are
currently incapable of accelerating electrons beyond a limit of 20-100
TeV (for Cassiopeia A the limit is 80
TeV)\footnote{\citet{Reynolds1999} found that Kes 73 could obtain an
anomalously high energy of 300 TeV. However, it has since been shown
that Kes 73 contains a pulsar, i.e. an entirely different source of
synchrotron emission.}. This work was repeated for 11 SNRs in the
Large Magellanic Cloud \citep{Hendrick2001} this time fitting a Sedov
model \citep{Borkowski2001} and \srcut\/ simultaneously, with similar
limits placed on the ability of SNRs to accelerate electrons. These
results cast some suspicion on the role of SNRs in accelerating ions to
energies even below the spectral steepening
at 10$^{15}$ eV in the integrated cosmic-ray spectrum.

The next step was to move from setting limits to actually describing
the synchrotron spectrum in SNRs with suspected X-ray synchrotron
emission.  \citet{Reynolds1998} showed that the maximum energy
attained by electrons from the shock acceleration process could be
limited by several different mechanisms: 1) electrons above some
energy $E_{max}$ could escape from the remnant, e.g. due to a lack of
MHD waves of the appropriate scale for scattering, 2) the remnant
could be young enough (or small enough) that there has not been
sufficient time to accelerate electrons beyond some $E_{max}$, or 3)
$E_{max}$ could represent the energy at which radiative losses
precisely balance gains. The lowest of the energy limits determines
the nature of the spectrum. These limits can be calculated using the
expressions given in \citet{Reynolds1998}.


For SN~1006, models limited by radiative losses were ruled out even by
pre-\asca\/ data \citep{Reynolds1996}. An estimate of the magnetic field
of order 4 \mg\/ (from TeV $\gamma$-rays assuming an inverse-Compton
origin; \citealp{Tanimori1998}), eliminated the age-limited model (the
magnetic field would have to be below 0.6 \mg\/ for the age limit to be
low enough to avoid exceeding the observed X-rays).  It should be
noted that the escape model is particularly simple to implement.
Unlike the loss and age limited models, it is approximately a
single-parameter model easily adapted for inclusion in X-ray spectral
analysis software.


In this paper we outline the theory behind the \sresc\/ model in \S
\ref{description} and then discuss the \sresc\/ submodels. We discuss
issues related to X-ray and radio observations in \S \ref{X-ray} \& \S
\ref{radio}. In \S \ref{fits6} we apply the new models to regions of
SN~1006. In \S \ref{Discussion} we discuss the results of our fits
including abundance information, and discuss the implications of those
results.
We then summarize our conclusions in \S \ref{conclusions}.

\section{Description of models} \label{description}

\subsection{Synchrotron model} \label{synch}

The synchrotron models used in this paper consider that the shock
everywhere accelerates a power-law distribution of electrons, with an
exponential cutoff above an energy \emax\/ whose value depends on
various physical parameters: remnant age, shock obliquity angle, shock
speed, and magnetic field strength \citep[details may be found in][]{Reynolds1998}. The models then evolve that distribution behind
the shock including adiabatic and radiative losses, and calculate the
volume emissivity of synchrotron radiation at each point in the
remnant by integrating that distribution over the single-electron
synchrotron emissivity. Images and total flux spectra are obtained by
appropriate integrations over the volume emissivity.  All synchrotron
models, including \sresc, \cp, \limb, and \ctr\/, begin with the
assumption of Sedov dynamics and a power-law distribution of electrons
up to a maximum energy $E_{max}$ which in general may vary with both
physical location within the remnant and with time.


As discussed in \S \ref{context}, we have excluded age and loss limited models,
in favor of the escape model.
In this
model electrons are presumed to escape upstream, probably due to an
absence of MHD waves beyond a certain wavelength
$\lambda_{\rm max}$.
The wavelength $\lambda_{\rm max}$ corresponds to an energy \emax\/  of
electrons. Electrons with gyroradius $r_g$ scatter resonantly with waves of
wavelengths $\lambda = 2 \pi r_g \cos \psi = 2\pi(E/eB) \cos \psi,$
where $\psi$ is the electron pitch angle. Therefore electrons will
escape upstream once their energy reaches an \emax\/  given by

$$E_{\rm max} = \lambda_{\rm max} \ e \ B_1/4 = 12.0 \  
\lambda_{17}
\ B_{1\mu G} \ \  {\rm erg},$$
where $B_{1\mu G}$ is the upstream magnetic-field strength in
units of \mg, $\lambda_{17}$ is $\lambda_{\rm max}$ in 
units of $10^{17} {\rm cm}$, and we have averaged over pitch angles.

We presume initially that
$\lambda_{\rm max}$ and $B_1$ are uniform outside the remnant.
Downstream, these electrons radiate in a magnetic field $B_2 \equiv
r_B B_1$, producing photons primarily at a frequency $\nu_{\rm max}
\propto E_{\rm max}^2 B_2,$ more precisely
$$\nu_{\rm max} = 1.05 \times 10^{15} \lambda_{\rm max}^2 B_{1\mu G}^3
\left(r_B \over 4\right) \ {\rm Hz}$$
with the magnetic compression ratio $r_B$ varying from 1, where the shock
normal is parallel to the upstream magnetic field (``pole''), to the
full compression ratio, assumed to be 4, where the shock is
perpendicular (equatorial ``belt'').  So the total spectrum involves a
superposition of spectra with different turnover frequencies over a
range of about 4; a fit to an observed spectrum does not produce a
unique \emax. 
This situation is in contrast to the much
simpler XSPEC model \srcut, where the spectrum is a power-law
with an exponential cutoff at some \emax -- presumed to be
the same everywhere in the source.

The implementation in XSPEC of the escape-limited synchrotron model
\sresc\/ has three parameters:  

\begin{enumerate} 

\item the radio flux measurement in Janskys at 1 GHz (\norm)

\item $\alpha$, the radio spectral index, where flux density $\propto \nu^{-\alpha}$  

\item a characteristic rolloff frequency, fitted by XSPEC, in Hz (\rolloff\/)  \end{enumerate}

For historical reasons, the frequency \rolloff\/ fitted
by the XSPEC implementation related to  $\nu_{\rm max}$ above by 

$$\nu_{\rm rolloff}  = 5.3\ \nu_{\rm max}.$$

At this frequency the spectrum is about a factor of 6 below a power
law extrapolation from lower frequencies.  Note that fixing
\rolloff\/ does not determine $\lambda_{max}$, $B_{1}$ or $r_B$
independently but only the combination $\lambda_{max}^2 B^3_{1} r_B$.


\begin{figure}
\plottwo{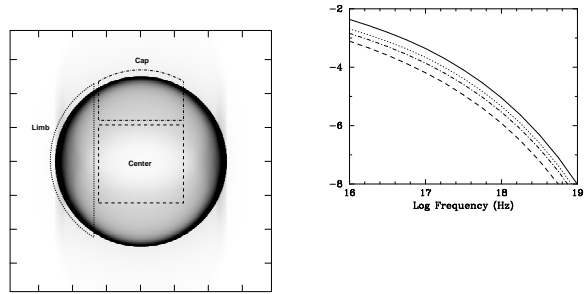}{f1b.eps}
\caption{Model and spectrum for $\phi = 30^\circ$. Dashed line is
the center region, dotted line is the limb regions and dot-dashed line
is the cap regions. Solid line is the total of all regions.} 
\label{fluxtot30}
\end{figure}

\begin{figure}
\plottwo{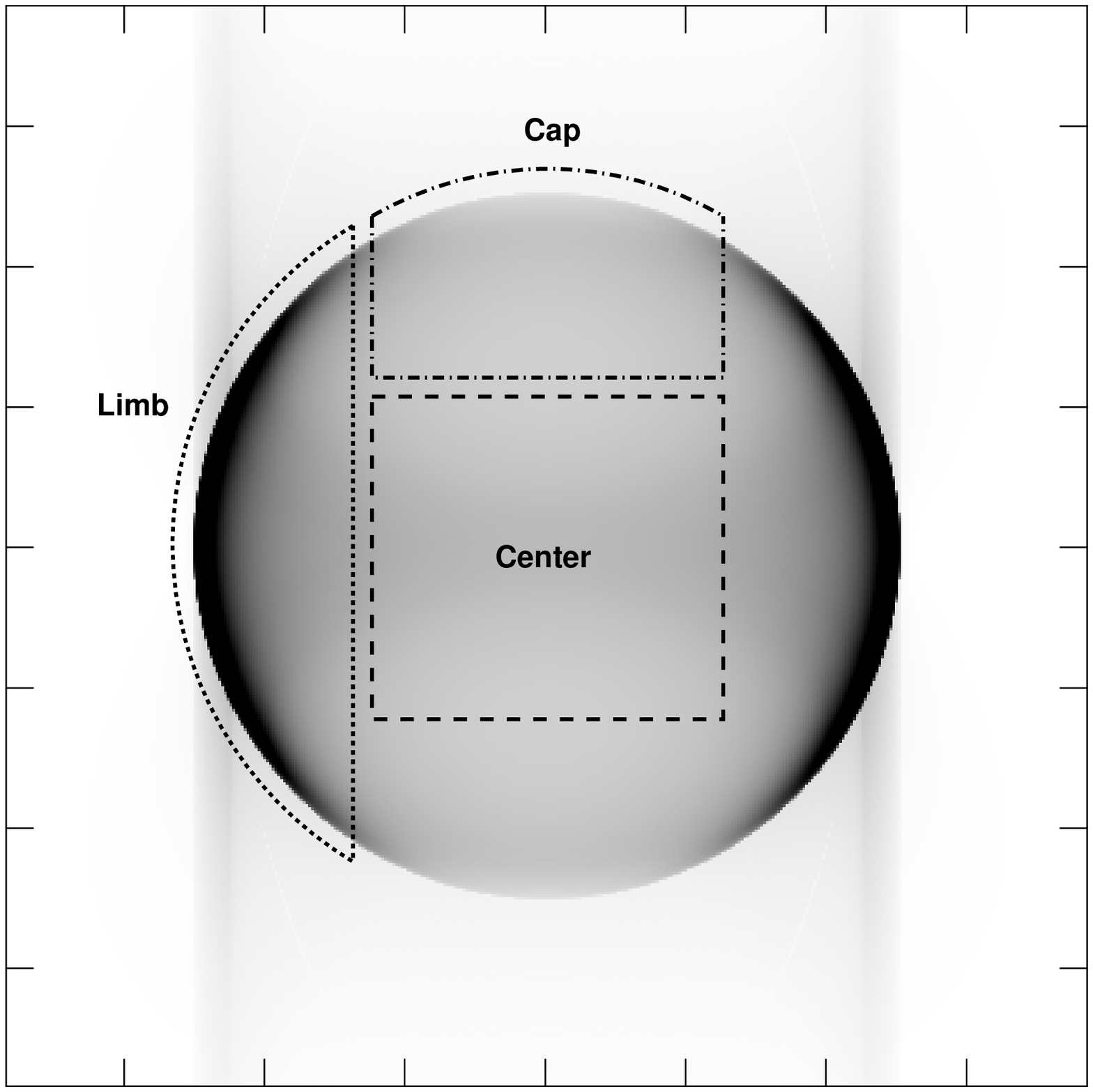}{f2b.eps}
\caption{Model and spectrum for for $\phi = 60^\circ$. Dashed line is
the center region, dotted line is the limb regions and dot-dashed line
is the cap regions. Solid line is the total of all regions.} 
\label{fluxtot60}
\end{figure}

\begin{figure}
\plottwo{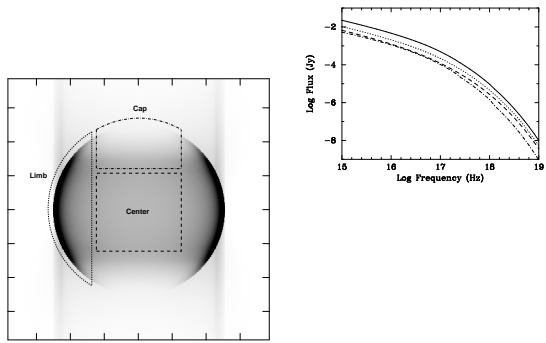}{f3b.eps}
\caption{Model and spectrum for for $\phi = 90^\circ$. Dashed line is
the center region, dotted line is the limb regions and dot-dashed line
is the cap regions. Solid line is the total of all regions.} 
\label{fluxtot90}
\end{figure}

The submodel spectra are plotted in
Figures~\ref{fluxtot30}-\ref{fluxtot90}, for three values of the aspect
angle $\phi$ between $B_1$ and the line of sight.  The
integrated spectrum is insensitive to $\phi$, since at these
frequencies the emission is extremely optically thin, but the
submodels differ since the flux is apportioned differently among the
regions as a function of $\phi$.  For instance, the distinction
between ``cap'' and ``limb'' begins to vanish as $\phi$ decreases; at
$\phi=0$ (not shown), the image would be circularly symmetric on the
sky (though our definitions of {\it cap} and {\it limb} encompass
different fractions of the image area).  These are spectra for a set
of parameters $B_1$, etc., whose values are arbitrary, for the purpose
of comparison among subregions.  For all values of $\phi$, we notice
that even though the limb regions cover the smallest area, the limbs still
make the largest contributions to the total.  At low enough
frequencies where the curvature has not yet become important, the limb flux
is about half the total for all values of $\phi$, while the center
contribution to the total drops from 0.32 to 0.19 (as $\phi$ drops
from $90^\circ$) and the cap rises from 0.21 to 0.34.

The limb emission is dominated by lines of sight nearly tangent to the
shell at the very edge, and as the frequency increases and the
emission is restricted (by electron energy losses) to thinner and
thinner postshock regions, those lines of sight shorten quickly, so
that for larger values of $\phi$, the integrated limb spectrum falls
off more quickly than that of the center, which for larger values of
$\phi$ also includes the highest-energy electrons (found near the
``equator'').  The softest spectrum is that of the caps, which include
regions where the obliquity $\theta_{\rm Bn}$ is always near 0
(parallel shocks); for low values of $\phi$, the center does not
contain any of the ``equator'' regions so its spectrum is softer as
well.  All these differences in the rolloff frequencies (the frequency
where the decrement from a power law is approximately a factor of 6) amount to
no more than 50\% of the input model value of $\nu_{\rm rolloff}$
because of the relatively steep dropoff of the spectrum.

Anticipating the application of these models to SN~1006, we 
note that few objects are likely to be so symmetric that
all parts of these models will fit the respective regions of
the remnants.  In the \srcut\/ and spatially-integrated \sresc\/ 
models already in XSPEC, the models are given as decrements
below the extrapolation of the power-law from radio frequencies.
The principal
difference among the region submodels is in normalization; the 
amount of curvature difference is relatively small, largest at
$\phi = 90^\circ$ and decreasing as $\phi$ decreases.

\begin{figure}
\plotone{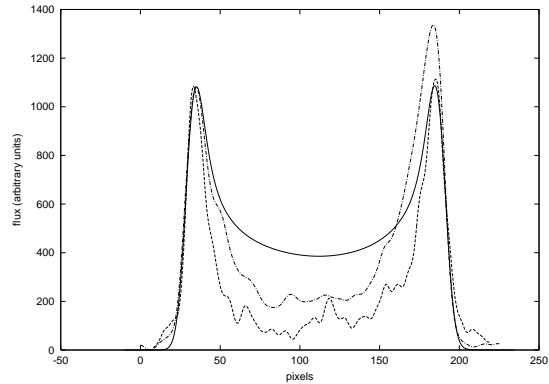}
\caption{A comparison of a limb-brightened model (solid line) to
actual NE-SW slices from the Parkes+Molonglo radio image (dash-dot) and the
\asca\/ \gis\/ image (dashes). Note that there is not enough central flux in
either the X-ray or radio image for the morphology of SN~1006 to be
explained entirely by limb brightening.} 
\label{slices}
\end{figure}

In applying these models to SN~1006, we are forced to confront a
significant problem: as demonstrated in Figure \ref{slices} at
both radio and X-ray wavelengths
the model overpredicts the emission from the center of SN~1006. 
 We will
discuss SN~1006's lack of symmetry briefly in Section
\ref{morphology}.
However, we do not attempt to model the detailed radio emission in this
paper.
We can model X-ray emission without presuming a
radio structure by using the observed radio fluxes and decrements in
our XSPEC models.  That is, we measure the dropoff in the X-ray
spectrum from the observed, not the theoretical, radio fluxes.

Similarly, we assume that the same value
of $\lambda_{\rm max}$ characterizes the emission in all regions. If that is
not the case, one should in principle produce a totally new
theoretical model incorporating some assumed spatial variation
of $\lambda_{\rm max}$.  However, given the relatively small variations in
curvature for different regions, we shall adopt a more phenomenological
approach, and allow the fitted rolloff frequencies in the different
regions to vary if necessary.

We shall proceed under the assumption that the bilateral symmetry of
SN~1006 is due to an ambient magnetic field close to the plane of
the sky, that is, $\phi \cong 90^\circ$.  The differences of model
curvatures with obliquity in the \asca\/ band are small enough
that this should not be a restrictive assumption.  
In general, 
substantially longer integration times would be necessary for
current X-ray satellites to detect these subtle differences in 
curvature.

\subsection{Thermal model}

Since our primary interest is in the nonthermal emission, we fit
thermal emission from the remnant with the simplest reasonable model
\vpshock. Wherever we specify that a thermal model was used, we mean
\vpshock. This is a plane-parallel shock model with variable
abundances \citep{Borkowski2001}. It represents an improvement over
single-temperature, single-ionization timescale, non-equilibrium
ionization models by allowing a distribution of ionization
timescales. 
In principle one might expect improvements in the results of thermal
modeling through the use of models at the next level of sophistication,
such as Sedov or multitemperature models. However, it is already clear from
high quality \xmm\/ and \chandra\/ observations that thermal spectra demand an
even higher level of modeling. A promising method is demonstrated in
\citet{Badenes2003} where one-dimensional simulations of different types
of explosions are used to generate synthetic spectra which are then
compared to X-ray observations of Tycho. We will limit ourselves to
the simplest reasonable thermal models, which here play a subordinate role
to the nonthermal models.


\section{X-ray observations}
\label{X-ray}

Unless stated otherwise \sis~0 \& 1 datasets from the same observation
were fit for each region. Full descriptions of the datasets used are
given in Table \ref{data6}. We used \sis\/  BRIGHT data at high and
medium bit rate.  The standard REV2 screening was used, and data were
grouped with minimum of 20 counts per channel for valid $\chi^2$
analysis.  The background spectra were obtained from the 1994 Blank
Sky event files in the case of PV 4-chip observations and from off
source areas for AO4 2-chip observations. We compared the results of
subtracting the two types of backgrounds and found the differences in
the resulting data were minor. What was not minor were gain shifts
between GIS and SIS data. We intended to fit GIS and SIS data
simultaneously to obtain better statistics. However, we found
significant line shifts between the GIS and SIS.
The difference was 70-80 eV in strong line energies (silicon is
prominent). Experimental GIS matrices reduced the shift in Si line
energy by $\sim$20 eV, but they were unable to completely remove the
shift. We therefore chose the SIS instrument over the GIS for energy
reliability and higher spatial and spectral resolution.  

Spectra from the \gis\/  do have one advantage over the
\sis\/ spectra. The \gis\/ detectors have a higher effective area at
high energy -- since this extra signal to noise could have
implications for the regions dominated by nonthermal emission we
verified the results obtained from the \sis\/ spectra by checking 
elliptical \gis\/ regions in the NE and southwest (SW). The energy shifts
between \gis\/ and \sis\/ are less important for measuring the
continuum.  The \gis~2 and \gis~3 data were jointly fit for each
limb.  We obtained radio fluxes (\norm\/) from the same regions used
to extract X-ray counts, and obtained background from regions of
similar size outside each respective limb. We discuss the results of
this comparison at the end of Section \ref{limbs}.






\section{Radio observations}
\label{radio}

In Paper I, since the entire remnant was being fit, we obtained the
normalization for the synchrotron models from single dish flux
measurements. However, in order to separate limb from center emission
we now needed accurate spatially resolved fluxes.
Interferometric images have the fundamental problem
that zero-spacing flux (the total flux in the image) is not measured. It is
well known that maps without this ``zero-spacing information'' are
missing the information needed to accurately reconstruct the total 
flux \citep{Holdaway}.
The absence of short interferometric spacings showed
up prominently for SN~1006 as negative flux densities in the center,
NW, and SE regions in the images published in
\citet{Reynolds1986,Reynolds1993} and \citet{Moffett}. 
``Zero-spacing information'' can only be restored to interferometric
maps by adding single dish observations. Therefore we obtained an 843
MHz map created with data from both the Molonglo Observatory Synthesis
Telescope (MOST), an East-West parabolic interferometer with UV
coverage from 15 meters to 1.6 kilometers, and the Parkes 64-meter
radio telescope, resulting in an image with resolution of $44'' \times 66''$ 
\citep{Roger}.
The total flux in the Roger et
al. image agrees with single dish measurements and in this map no
regions of the SNR have negative fluxes.

We measured the radio flux in the exact regions from which 
we extracted spectra
by using a region file in sky coordinates to extract both the spectrum
and an X-ray image. That image was loaded into AIPS, convolved to a
resolution of 11\arcsec\/ to smooth the X-ray sampling, and aligned to
the radio image with AIPS task {{\it HGEOM}. The radio image was then
clipped everywhere the X-ray image was blank, and the remaining flux
was measured at the observing frequency (0.834 GHz) and then
extrapolated to 1 GHz according to the single dish spectral index of
0.6 \citep{Green}.

\section{Fits to data}
\label{fits6}

We have fit and plotted the \asca\/ data from 0.4-10.0 keV. While 0.4
keV is below where \asca\/ data are normally considered reliable
(generally above 0.6 keV and for the most recent data, only above 1.0
keV) we have chosen to use these data for several reasons. First, most
of the datasets used are from the early performance verification phase
(PV), prior to significant chip damage. Second, \sis0 and \sis1 are
still in good agreement in that region -- one test for data
reliability. Finally, there were additional problems fitting regions
from the center, southeast (SE) and northwest (NW) in the vicinity of 0.6
keV. Without including data from 0.4-0.6 keV, XSPEC was free to fit
models with extremely high flux at low energies.  While we do not
trust the data sufficiently to report abundances measured at low
energies, we believe the data can be trusted to rule out fluxes high
by a factor of two or more. Finally, the lowest energies in each fit
were dominated by thermal emission. As stated above, the main purpose
of this work was to test models against nonthermal emission, which
dominates at higher energies.
 
There were slight variations reported by \sis0 and \sis1 in the
measured flux. 
While \csn\/ could be slightly improved by allowing the normalizations
of the models in the two detectors to vary slightly, we did not choose
to do so, preferring a slightly higher \csn\/ over introducing further
uncertainties in \norm.

We began by fitting solar-abundance thermal models (\pshock) to each
of the regions (with a \limb\/ model in the NE and SW). These models fit
very poorly, with \csn\/ ranging from 3.6-20 and obvious line
residuals. We can say with certainty that even in the bright
limbs where the nonthermal emission dominates, SN~1006 is not well
described by solar abundance models. Henceforth, as we discuss our
fits, all the \vpshock\/ models we employ allow the individual
abundances to vary.


Table \ref{everything} contains the results of the fits to each region
of the remnant, for each model combination as described
below. Included are $\chi^2$ and the number of degrees of freedom
(DOF, column 1). The following two columns determine the nonthermal
model: the 1~GHz norm (the flux in Jy, measured from the radio map) in
column 3 (difference between \sis~0 and \sis~1 are a result of gap
location, rather than chip differences), and in column 4, the
\rolloff\/  ($\nu_{\rm rolloff}$ in Hz for all \sresc\/ submodels).
Thermal parameters from the \vpshock\/ model include: the temperature
found by each thermal model in units of keV (column 5) and the
ionization timescale $\tau \equiv n_e t$ in units of s~cm$^{-3}$
(column 6).  In column 7 we give the emission measure (\norm),
defined as
$\slantfrac{10^{-14}}{4\pi D^2} \int n_e n_H dV$ where $D$ is the
distance to the source in centimeters. The fitted abundances are
listed in column 8-12, given as $\case{<X/X_\sun>}{<Si/Si_\sun>}$
\citep[by number, not mass, from][]{Grevesse}. Since the hydrogen
abundance is poorly constrained by the fits, abundances were fit
relative to silicon. Errors given in Table \ref{everything} are 1.5$\sigma$ errors (87\%
confidence interval). As in
Paper I, for both \gis\/ and \sis\/ data, we assumed an absorption column density of 5$\times 10^{20}$
cm$^{-2}$, using the Wisconsin absorption model, {\it wabs}.

Note that, except where specified, all synchrotron models were fit
with normalization and spectral index fixed, as measured from radio
observations. There is considerable degeneracy in the model between
\norm\/, $\alpha$, and \rolloff, as demonstrated in Table
\ref{limits}, but radio measurements allow us to input, rather than
fit, \norm\/ \& $\alpha$.

\subsection{Brighter limbs: northeast and southwest}
\label{limbs}


\begin{figure} 
\plottwo{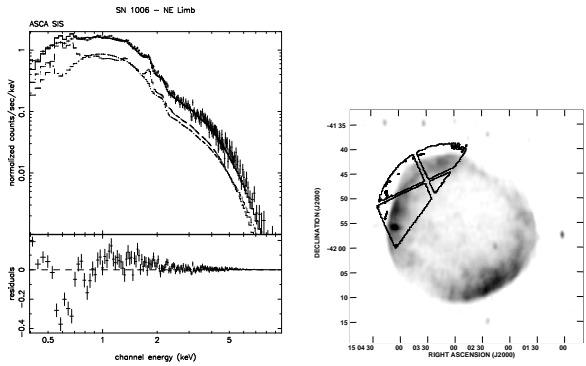}{f5b.eps}
\caption[Fit thermal: limbs]{Thermal and nonthermal fits to the northeast limb.
\label{NLboth}} 
\end{figure}

\begin{figure} 
\plottwo{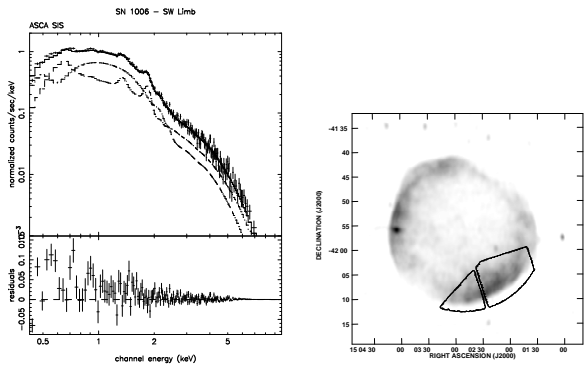}{f6b.eps}
\caption[Fit thermal: limbs]{Thermal and nonthermal fits to the southwest limb.
\label{SLboth}} 
\end{figure}

We began by searching for firm upper bounds to the synchrotron
emission from the bright limb regions, shown in Figures \ref{NLboth}
\& \ref{SLboth}.   We did this
by finding the maximum amount of the \limb\/ model tolerated by
the \sis\/ NE and SW regions, assuming a spectral index of 0.60
and \limb\/ \norm\/ measured from the radio, i.e. we raised
\mbox{\rolloff} until
it threatened to exceed the data at high energies. We found that the
spectra of the NE and SW regions are not identical (see
Figure \ref{maxsync}).
The NE could tolerate a maximum
\rolloff\/ of 7$\times10^{17}$ Hz while the SW could only
tolerate a maximum \rolloff\/ of 4$\times10^{17}$ Hz.

\begin{figure} 
\plottwo{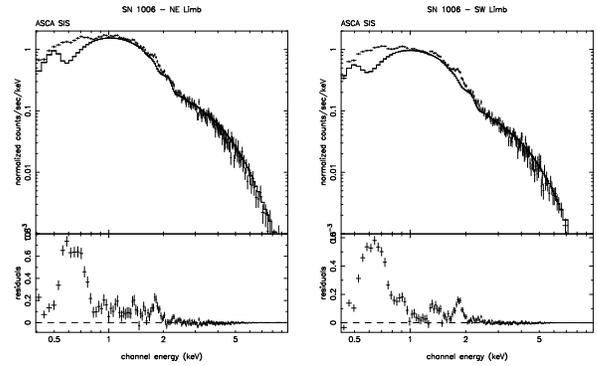}{f7b.eps}
\caption[Maximum nonthermal: limbs]{The northeast and southwest limbs
 with the maximum amount of nonthermal synchrotron plotted.
\label{maxsync}} 
\end{figure}

We then fit the data with combined thermal and nonthermal models,
using \mbox{\vpshock} (variable abundances)+\limb, 
shown in Figure \ref{NLboth}. The results for
the fit are given in Table \ref{everything}, row 1. The combined
thermal and nonthermal fit achieved a \csn $\sim$1.8. Residuals from
the fits primarily show discrepancies between \sis0 and \sis1 between
0.5 and 1.0 keV.

With the addition of a thermal model the \rolloff\/ for the NE
dropped to 3.3$\times10^{17}$~Hz with 1.5$\sigma$ limits of
2.6$\times10^{17}$ and 3.8$\times10^{17}$~Hz. The SW had a
\rolloff\/ of 2.3$\times 10^{17}$~Hz with 1.5$\sigma$ limits of
1.9$\times10^{17}$ and 2.6$\times10^{17}$~Hz.  
These fitted values of \rolloff\/ do appear significantly different,
though not by as much as the maximum values above.
This agrees with the higher radio-to-X-ray
ratio in the SE shown in Figure \ref{slices}. In order for the
radio-to-X-ray ratio to be lower than in the NE, \rolloff\/ must be lower
(since $\alpha$ is fixed.)
The thermal component of the NE
had a temperature of 1.85$_{1.81}^{1.89}$~keV
and ionization timescale of 5.2$^{6.8}_{3.8}\times
10^{9}$~s~cm$^{-3}$. The total flux from 0.4 to 10.0 keV in the NE
 was 6.1$\times 10^{-11}$~\egcms, of which 47\% was from the
synchrotron model. The SW had a temperature of
1.87$_{1.79}^{1.89}$~keV and an ionization timescale of
1.5$_{1.4}^{1.7}\times10^{10}$\cms. The total flux in the SW was
4.3$\times 10^{-11}$~\egcms, of which 46\% was from the synchrotron
model.


The \gis\/ data, in contrast to the \sis\/ data, show no obvious lines
between 2 and 10 keV. Therefore there is little purchase for a
thermal+nonthermal fit, although for comparison with \sis\/ results we
did try fits with \sresc\/ and \sresc\/+\vpshock\/ models. Since the
\gis\/ has a much wider bandpass than the \sis\/ we also allowed the 
spectral index to vary, in order to eliminate the possibility that
freezing $\alpha$ within the narrower \sis\/ bandpass could introduce
false results in \rolloff\/ (as discussed above and shown in Table
\ref{limits} the degeneracy in \norm\/, \rolloff\/, and $\alpha$
decreases the usefulness of fitting $\alpha$).

The 1 GHz {\it norms} from the \gis\/ regions were 3.37 Jy for the NE
and 5.11 Jy for the SW. (All errors on \gis\/ observations are 1.65
$\sigma$, i.e. 90\% confidence interval.)
With $\alpha$ fixed to 0.6, we obtained values of 
\rolloff\/ of $5.89_{5.79}^{5.99}\times 10^{17}$ Hz in the NE and $\nu_{\rm rolloff} =
3.25_{3.19}^{3.30} \times 10^{17}$ Hz in the SW. \
With $\alpha$ allowed to vary we obtained $\nu_{\rm rolloff} =
2.61_{2.19}^{2.95} \times 10^{17}$ Hz and $\alpha = 0.55 \pm 0.01$ in
the NE, and $\nu_{\rm rolloff} = 2.62_{2.25}^{3.12}
\times 10^{17}$ Hz and $\alpha = 0.59 \pm 0.01$ in the SW.
The thermal+nonthermal fits were carried out with \sresc\/ plus a solar
abundance \vpshock\/ model.
With $\alpha$ fixed at 0.6, 
we found \rolloff\/ to be $4.50_{3.57}^{5.11}
\times 10^{17}$ Hz in the NE and $2.65_{1.95}^{3.08} \times 10^{17}$ Hz
in the SW. 

This confirms results found from the \sis\/ fits, including
the slightly higher rolloff frequency in the NE.  In the
thermal+nonthermal fits the \gis\/ do not show any significant
differences in \rolloff\/ that might be expected from the better
\gis\/ sensitivity at high energies. The results are consistent with
the results for \sis\/ fits, which are preferred since they more
accurately constrain the thermal emission.

\subsection{Fainter limbs: northwest and southeast}
\subsubsection{Northwest}
\label{northwest}

Our initial assumption in applying the simple escape model was that
all regions of the remnant would be fit with varying amounts of \norm,
measured in radio observations, but the same spectral index and
\rolloff. We note that this procedure normalizes away any departure
of the {\bf radio} morphology from model predictions (for example,
the simple \sresc\/ model predicts higher brightness in the remnant
center than is observed).  It simply requires that no {\it additional}
departures from model predictions would be necessary between
radio and X-ray wavelengths. 

However, it is immediately clear that while the NE and SW regions have
similar \rolloff's (average 2.8$\times10^{17}$ Hz), that this value of
\rolloff\/ in the \cp\/ submodel, with \norm\/ appropriate to the NW
and SE, clearly contradicts the NW and SE data, even {\bf before} addition
of a thermal model required by the presence of lines (see Figures
\ref{npc}b and \ref{spc}b). This agrees with the high spatial
resolution \xmm\/ results of
\citet{Decourchelle2002} which found \rolloff\/ varying with both radius
and azimuth. 

\begin{figure}
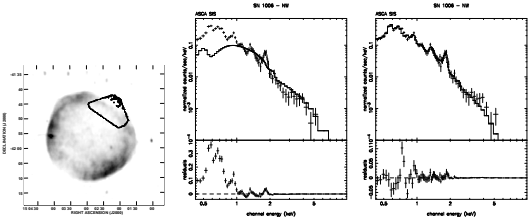

\epsscale{.30}
\plotone{f8a.eps}
\plotone{f8b.eps}
\plotone{f8c.eps}
\caption[all NW fits]{Northwest:
a) \sis~region shown on the Parkes radio image
b) high \cp~model with \rolloff~measured at the limbs
c) thermal model \vpshock.
\label{npc}}
\end{figure}

\begin{figure}
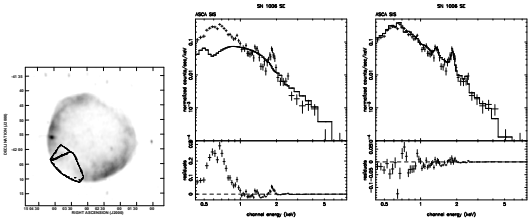

\epsscale{.30}
\plotone{f9a.eps}
\plotone{f9b.eps}
\plotone{f9c.eps}
\caption[all SE fits]{Southeast:
a) \sis~region shown on the Parkes radio image
b) \cp~model emission predicted by the \rolloff~measured at the limbs
c) thermal model \vpshock.
\label{spc}}
\end{figure}

Fits to the NW region with a thermal-only variable abundance model
(shown in Figure
\ref{npc}c) yielded a $\chi^2_\nu$ of 1.80. Deficiencies in the thermal
model can be noted in residuals near 0.7 and 0.9~keV (Fe
L-shell or Ne K$\alpha$). Full results are given in Table \ref{everything}, row 5. The
temperature given by the fit was 1.2$^{1.3}_{1.0}$~keV and
$\tau=6.5^{7.9}_{5.5}\times10^{9}$~\cms. The total flux from this fit was
1.3$\times10^{-11}$~\egcms. While a fit with thermal+nonthermal models was
attempted the data preferred no nonthermal component.

It was critical to place upper limits on the amount of nonthermal
emission even in regions where no nonthermal component was required so
we used the following procedure: we added a nonthermal model
(beginning with $\alpha=0.60$, \rolloff=2.8$\times 10^{17}$ Hz and
norms measured from the radio image) to the best thermal-only fit
(given in Table \ref{everything}, row 5) and then changed each of the
three nonthermal parameters in turn ($\alpha$, \rolloff, and \norm),
holding the others fixed until the $\chi^2$\/ rose by $\sim$2.3, a 1.5$\sigma$
change. The results of these fits are given in Table \ref{limits} with
the varying quantity in bold face. With this method we found that for
the NW data the \cp\/ model had a maximum \rolloff\/ of
1.1$\times10^{17}$~Hz, or a maximum \norm\/  of 0.25 Jy at 1 GHz, or a
spectral index $\alpha$ of 0.67 or steeper. 
This places a limit of $\leq$8\% on
the nonthermal contribution to the NW flux.

\subsubsection{Southeast}

As with the NW, using the value of \rolloff\/ measured in the NE
and SW regions overpredicted the observed flux in the SE (see
Figure \ref{spc}b). Thermal fits with variable abundances
to the SE (see Figure
\ref{spc}d) were significantly worse than the NW. \csn\/ was
$\sim$3 for a single \vpshock\/ model, a two-\vpshock\/ model (not shown), and
a \cp+\vpshock\/ model. Interestingly, despite a much worse \csn\/ the
residuals look very similar to those in the NE -- obvious
line-like residuals at 0.7 and 0.9~keV.

The thermal-only model (Table \ref{everything}, row 7) had a
temperature of 1.1$^{1.2}_{1.0}$~keV and a
$\tau=2.3^{2.8}_{1.8}\times10^{10}$~\cms. As above for the NW,
while a combined thermal and nonthermal model was tried, the data
preferred no nonthermal component at all.


Following the method outlined above for the NW, we set limits on the
nonthermal emission. In the SE we can exclude a \rolloff\/ above
1.1$\times10^{17}$ Hz or a spectral index, $\alpha$ flatter than 0.67, or a
\norm\/  above 0.2 Jy at 1 GHz. The total flux of this faint region is
8.6$\times10^{-12}$~\egcms\/ and the maximum nonthermal contribution is $\leq
9\%$.

\subsection{Center}

\begin{figure}
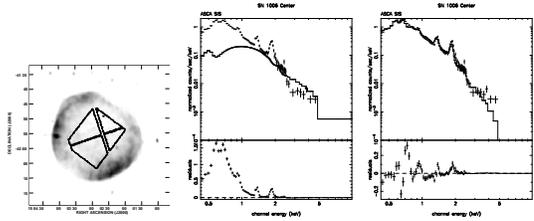

\epsscale{.30}
\plotone{f10a.eps}
\plotone{f10b.eps}
\plotone{f10c.eps}
\caption[all NW fits]{Center:
a) \sis~region shown on the Parkes radio image 
b) \ctr~model emission predicted by the \rolloff~measured at the limbs
c) thermal model \vpshock.
\label{center}}
\end{figure}

We turn now to the center of SN~1006. Unlike the NW and
SE, the center data could just tolerate the nonthermal model
\ctr\/ with fixed \rolloff\/, shown in Figure \ref{center}b. However, joint fits
soon reduced the nonthermal component to $\leq$5\% (using the method
outlined in fits for the NW and SE).  Both combined and thermal-only
models had very high \csn ($>5$).  Thermal-only fits (Figure
\ref{center}d and Table \ref{everything}, row 3) reported a
temperature of 0.82$^{0.85}_{0.76}$~keV and a
$\tau=2.5^{2.8}_{2.3}\times10^{10}$~\cms\/. 
As with the NW and SE, when a thermal+nonthermal model
fit was attempted the data preferred no nonthermal component.
All fits to the center show residuals
similar to the NW and SE.  


The thermal fits for each region of the SNR reveal elevated
abundances. 
We can make a separate case for supersolar abundances by
measuring the equivalent width of the well-separated Si K$\alpha$
line in the center where the lines
are strongest. The result is an equivalent width for silicon of 600
$\pm$ 50 eV (1.65 $\sigma$, 90\% confidence). We draw conclusions from this in Section \ref{eqw}.

\section{Discussion}
\label{Discussion}

\subsection{Difference between the NE and SW limbs}

The NE and SW regions do have slightly different spectra, with values
of \rolloff\/ differing by a significant amount in both \sis\/ and
\gis\/ fits including a thermal component.
(The \rolloff\/ values bracket the Paper I value averaged over the
remnant of $3.1 \times 10^{17}$ Hz.)  This is not, however, sufficient
to explain the differences in TeV detection of the NE and SW regions
(assuming as in Paper I that the TeV detection is cosmic microwave
background photons, inverse-Compton scattered by the same population of
electrons that produce the X-ray synchrotron emission). The small
difference in \rolloff\/ between NE and SW regions only produces a
difference of 5\% in this flux at 1 TeV. Even \rolloff's differing by
factor of three only produce a difference of $\sqrt 3 \sim 70$\% in
$E_{max}$.  A difference in the spectral index is far more likely to
produce the differences between synchrotron emission in the NE and SW
regions which would be required to explain differences in $\gamma$-ray
detection (although, as discussed below the spectral index cannot
account for the entire change).

Using the formula given in \S \ref{synch} and assuming a magnetic compression ratio
$r$=4 and value \mbox{$B_1$= 3 \mg} as found in Paper I, we find that the
\rolloff\/ of 3.3$^{3.8}_{2.6}\times10^{17}$ Hz in the NE gives a
$\lambda_{max}$=1.5$^{1.6}_{1.3}\times10^{17}$ cm and in the SW a
\rolloff\/ of 2.3$^{2.6}_{1.9}\times10^{17}$ Hz gives
$\lambda_{max}$=1.2$^{1.1}_{1.3}\times10^{17}$ cm.  These values
imply electron escape above 33 TeV in the NE or 28 TeV in the SW, 
similar to the value of 32 TeV quoted in Paper I, as is to be expected.


\subsection{Thermal fits}


As stated earlier, the focus of this work was on testing nonthermal
models.  Therefore only the simplest plane shock model was used to fit
thermal emission. We ruled out solar-abundance thermal emission in
each region of the SNR and then used the variable-abundance model,
\vpshock\/, for the remainder of our fits. 

The quality of the fits varied widely across the SNR. In bright regions
such as the NE, the SW, and the NW, the fits gave marginal
values of 
$\chi^2_\nu \sim 1.8$. Fainter regions such as the center and the
SE were poorly fit by
\vpshock\/ with $\chi^2_\nu > 5.0$. However, despite different values of
$\chi^2_\nu$, the residuals of all regions with significant thermal
emission (SE, NW and center) had line-like residuals at 0.7 and 0.9
keV. This casts suspicion on even the better fits and suggests
problems with the underlying atomic data since this is where 3s to 2p
transitions in Ne-like Fe ions are found. These transitions appear to
be stronger than expected in astrophysical sources, including SNRs \citep[as
seen in][]{Heyden2002}.

In the current NEI v1.1 models in XSPEC which we use, Fe L-shell data
are based on calculations by \citet{Liedahl1995}
for electron collisional excitation of Ne- to Li-like Fe ions, and on
older atomic data for inner-shell processes in Mg-like and Na-like Fe
ions \citep{Hamilton1984}. Indirect processes such as
inner-shell collisional ionization, radiative and dielectronic
recombination, and resonance excitation are not included, although
they have been shown to be important in enhancing 3s to 2p transitions
in the Ne-like ion in particular \citep{Gu2003}. The enhancement of 3s to 2p
transitions in the Ne-like ion due to the inner-shell ionization of
the Na-like ion might be particularly important in high-temperature
plasmas at low ionization ages, such as those encountered in SN
1006. In addition, \citet{Chen2003} showed recently
that atomic resonances in the electron collisional excitation of
Ne-like Fe are also important, although they have been neglected in
the past.  So there are many deficiencies in Fe L-shell atomic data
which could result in poor fits in the 0.7-0.9 keV range.

We tested this hypothesis by adding lines arising from the inner-shell
ionization of the Na-like Fe ion into XSPEC NEI v2.0 models. These
models are based on the Astrophysical Plasma Emission Database
\citep[APED;][]{Smith2001}, and are currently missing
several processes important in NEI conditions, but they allow for easy
modification of the atomic database. We added three lines of the
Ne-like ion at 17.098, 17.053, and 16.778 ${\rm \AA}$ (0.725, 0.727, and
0.739 keV, the strongest at 0.725 keV). For a shock model with
parameters appropriate for the NW region of SN~1006 ($kT=1.17$ keV and
$\tau = 6.5\times 10^{9} {\rm cm}^{-3}$), their contribution to the
Fe spectrum is substantial. So it appears certain that the current
fits to SN~1006 spectra use inadequate Fe L-shell atomic data, and
that one problem with current thermal fits can be traced to Fe
L-shell data. In particular, missing lines from inner-shell ionization
of Na-like Fe ions are certainly a problem for the NW, and most likely
elsewhere. Since this paper is primarily concerned with the nonthermal
emission and better spectral information (for smaller spatial scales)
has been obtained by
\xmm\/ and \chandra\/ we defer advanced models to a future paper.

This may not be the only issue preventing us from finding adequate
thermal fits. X-ray spectral modeling of Type Ia SNRs
\citep{Badenes2003} has revealed significant differences in physical
conditions between chemically-distinct layers of shocked ejecta.
In particular, the emission-measure averaged ionization timescale of Fe
appears to be shorter than that of other abundant elements in most models
considered by \citet[][Fig. 6]{Badenes2003}. It is then unlikely that
the chemically homogeneous \vpshock\/ model can provide a satisfactory
fit to both Fe L-shell lines and lines produced by other elements.
Substantial improvements in both Fe L-shell atomic data and in spectral
models are required in order to make further progress in modeling
SN ejecta in SN~1006. 

The Paper I fits to the total flux of SN~1006 achieved better values,
$\chi^2_\nu$=1.2.  However, the better fits are causes by two factors:
the poorer spectral resolution of the \asca\/ \gis\/
instrument and the fact that the integrated spectrum is dominated by the
synchrotron contributions of the two bright limbs which are well
described by our models.


One conclusion that can be drawn from the thermal fits in Table
\ref{everything} is that while the thermal fits may be oversimplified,
the \vpshock\/ model is clearly sensitive to both thermal parameters and
abundances. Whereas we might have expected large uncertainties, the
1.5$\sigma$ statistical errors are rarely more than 8\%.  Clearly,
the data are capable of constraining thermal models; the correct
thermal model should yield significant information.

\label{eqw}

While our thermal model fits to the center were of poor quality statistically,
they strongly implied elevated abundances of heavy elements such as Si.
Fortunately, a more model-independent criterion is available: the
equivalent width of the Si K$\alpha$ complex (Si XIII).  Our measured value
of $600 \pm 50$ eV is higher than a plane shock with solar abundances
could achieve until $\tau \sim 3 \times 10^{10}$ cm$^{-3}$ s, by
which time the shock has $kT$ considerably above 1~keV, and should
have much higher
ratios of  \ion{O}{7}
to 
\ion{Si}{13}
than we observe \citep{Long2003}.
We conclude that the central emission is dominated by shocked ejecta.

\subsection{Nonthermal results}

Our spatially resolved subsets of \sresc, \cp, \limb, and \ctr,
differ surprisingly little from one another or from the total
(\sresc\/ itself), though the differences we do see (\S \ref{synch}) are
in the sense one would expect.  However, for these \asca\/ data, and
in fact for most data from current X-ray telescopes, the differences
between the spatially integrated \sresc\/ model and the submodels are
not likely to be distinguishable.  This could not have been predicted
in advance, but it is fortunate for data analysis.  However, our
conclusions about the lack of cylindrical symmetry are stronger for
our having based them on the appropriate submodels that symmetry
would require.

The most striking result from the spatially resolved fit is that the
\sresc\/ submodels with values of \rolloff\/ that fit the limbs, do not
fit the spectrum in the center, NW and SE. There are
several possible explanations, with interesting consequences.  First,
it is possible that the \sresc\/ submodels are an accurate description
of the nonthermal emission and that the value of \rolloff\/  differs
from one region to the next. Since \rolloff\/~$\propto \lambda^2
B^3$ a difference in \rolloff\/ would imply a change in the external
magnetic field or the maximum wavelength of MHD waves available for
scattering. The variation of the \rolloff\/ parameter across the remnant
marks a significant departure from the simple assumptions of the
escape model.  We can offer no justification for this
but if $B$ or $\lambda_{max}$ could be isolated they would be
important probes for the shock physics.

As demonstrated in Table \ref{limits}, the \sresc\/ submodels
show some degeneracy between \rolloff\/  and the radio spectral
index $\alpha,$ in that a given X-ray to radio flux ratio can
be produced (within limits) by flatter spectra with lower 
\rolloff, or steeper spectra with higher \rolloff.
A large source of error in all synchrotron fits is the
uncertainty in $\alpha$. An uncertainty in $\alpha$ of $\pm$ 0.1 at 1
GHz can change the extrapolated flux in the 2 to 3 keV range by a
factor of 7. In addition to being poorly measured by single dish radio
observations, the spectral index could vary across the source. We can,
however, rule out the possibility that spectral index is {\bf solely}
responsible for the variation in \rolloff. While radio observations
cannot constrain $\alpha$ to better than $\pm$0.02, single-dish
observations firmly exclude an $\alpha$ as steep as 0.67.  Since the
two limbs dominate the total radio emission as they do the X-ray
emission, the kind of difference between limbs that would be required
seems unlikely. We cannot, of course, rule out a contrived
situation in which the spectrum of the SW suddenly steepens
significantly compared to the NE at radio frequencies higher than have
been measured.  Nevertheless, the question of a varying spectral index
across the remnant is intriguing. Many researchers have searched for
variations in radio spectral index in SNRs as an indication of changes
in shock acceleration. However, interferometric observations are ill
suited to this kind of study \citep[see discussion in][]{Dyer1999} and
current observations of SN~1006 can neither support nor contradict
this premise.

A second possibility is that the physics in the synchrotron model that
underlies \sresc\/ and submodels is inadequate to describe the
nonthermal emission.
One critical assumption is that the diffusion coefficient
$\kappa$ is proportional to the energy. A divergence from this assumption
at the highest energies could have serious consequences.

\subsection{Morphology}
\label{morphology}

We began our investigation by assuming cylindrical symmetry about the
NW-SE axis. We can now rule this out with a fair amount of certainty.
If SN~1006 were cylindrically symmetric a line of sight through the
center would sample a predictable amount of the nonthermal emission
which brightens the limbs. This is ruled out by the fact that the
\sresc\/ submodels over-predict the high energy flux in the NW, SE and
center, as demonstrated in Figures \ref{npc}b, \ref{spc}b and
\ref{center}b.  This deficit of central emission is in addition
to the deficit already present between observed and theoretical
radio images in the \sresc\/ model, since we are using measured
rather than theoretical values of \norm.  An explanation which might
account for the weaker central emission in radio compared to the
simple model (for instance a substantial radial component to the
magnetic field) cannot explain the additional deficit in X-rays.
Any explanation must impact the X-ray emission differently than the
radio emission.


That SN~1006 is not simply limb brightened due to a
NW-SE symmetry is an uncomfortable conclusion, since it implies a
preferred orientation between SN~1006 and the observer, although there
are other arguments for asymmetry from \citet{Willingale}, based on
ratios between the X-ray emission in limbs and center as observed by
\rosat, and arguments derived from the spectra of the
Schweizer-Middleditch star \citep{Hamilton1997}.

This asymmetry could take several forms. It is possible, likely in
fact, that the upstream magnetic field is not uniform. There
is already evidence for asymmetry in the NW, where
H$\alpha$ is strongest -- presumably the ambient medium outside the
NW is, for some reason, less ionized. This would have an impact
on electron acceleration.  This could affect $\lambda_{max}$ in
particular.  


It is possible, of course, that fundamental assumptions made in the
\sresc\/ model are incorrect.  While the agreement of the model with
the bright limbs argues that the basic picture of electron
acceleration to a power-law with an exponential cutoff is sound, at
least in those regions, it is possible that a completely different
geometry holds, for instance, one in which the bright limbs are
``caps'' seen edge-on rather than brightened limbs of an equatorial
``belt.'' The symmetry axis would then run NE-SW, but would need to be
nearly in the plane of the sky.  The reasons that the polar caps are
brighter in synchrotron radiation in such a picture would need to
involve some superior feature of parallel shocks for electron
injection or acceleration, or perhaps magnetic-field amplification as
suggested by \citet{Bell2001}.  A fundamental defect of such a picture
is that remnants for which the ambient magnetic field is closer to the
line of sight should appear in radio synchrotron images as two maxima
surrounded by a steep-spectrum halo, a morphology which is not seen in
any Galactic remnant.

\subsection{Comparison with {\it Chandra} Results}

\chandra\/ observed part of the NE limb in Cycle 1 and part of
the NW limb in Cycle 2.  The smaller field of view, and 
charge-transfer-inefficiency problems for the front-illuminated
CCDs, restricted initial analysis to much smaller regions than
those we analyzed here.  However, the results \citep{Long2003} are
consistent with ours.  In the NE, thin filaments appeared to have
completely line-free spectra, and were well described by an \sresc\/ model
with $\alpha = 0.54$ and \rolloff\/ of $6.9 \times 10^{17}$ Hz.
Since \chandra\/ was able to isolate a region immediately
behind the shock only a few tens of arcsec in size, it is to be
expected that the hardest electron spectrum (highest \rolloff) would
be found there.  Our \gis\/ value for \rolloff, with $\alpha$ allowed
to float, is close to this value, although slightly lower because of dilution
from thermal emission and nonthermal emission further downstream
where adiabatic and radiative losses begin to soften the spectrum.
A region several arcminutes behind the shock in the NE
was shown to have substantial line emission, and was fairly well
described by a \vpshock+\sresc\/ model with $kT \sim 1.4$ keV, $\tau \sim
3 \times 10^9$ cm$^{-3}$ s, and a \rolloff\/ of
$(3-6) \times 10^{17}$ Hz.  Elevated abundances of heavy elements
(chiefly Si and Fe) were required. These results are similar to 
our inferences about the thermal component in the NE, though these
of course apply to a much larger region than the {\it Chandra} results.
In the NW, a \vpshock\/ fit to a small region coincident with the 
H$\alpha$ optical filament gave $kT \sim 0.9$ keV, $\tau \sim 6 \times 10^9$
cm$^{-3}$ s, and required no nonthermal component (though it could
tolerate a weak one).  These results are comparable to those we
report here from a much larger region.  The broad similarities
between fitted values of small regions with {\it Chandra} data and 
larger averages with \asca\/ data give some confidence that the
systematic residuals we observe that degrade the statistical quality
of our fits are not spacecraft-dependent systematic effects.  (Of
course, they may be inherent in the models or atomic data, which 
were the same in the two analyses.)  While {\it Chandra} observations
did include regions closer to the center of SN~1006, those were on
ACIS-I chips subject to charge-transfer-inefficiency problems, and the data have not yet
been analyzed.

\subsection{Effect on Paper I Results}

These new investigations change few of the results in Paper I.
It is clear that synchrotron emission from the limbs dominates the total
fit: the value of \rolloff\/ in Paper I was $3.0_{2.8}^{3.1}\times10^{17}$~Hz, while fits to the limbs individually found $3.3\times10^{17}$ and
$2.3\times10^{17}$ Hz.

While in theory more precise values for the postshock magnetic field
strength, the energy in relativistic electrons and the electron efficiency
(9 \mg, 7$\times$10$^{48}$ ergs and 5\% in Paper I) could be obtained from
limb-only fits, in fact these will change insignificantly considering
other uncertainties.




 
\section{Conclusions}
\label{conclusions}

\subsection{Nonthermal Emission}

1.  From comparing the limb-to-center ratio in X-ray and radio wavelengths
(Figure \ref{slices}), we find that the morphology of SN~1006 is not solely
due to limb brightening of a cylindrically symmetric source about
a line parallel to the bright limbs.  This has the inescapable
consequence that the observer is now required to have a particular
vantage point with respect to the SNR.


2. \sresc\/ provides a useful framework to describe synchrotron
emission in SNR blast waves.
Spatially resolved subsets of the \sresc\/  model, \cp, \limb,
and \ctr, differ surprisingly little from the spatially integrated
model, with the greatest differences occurring when the upstream
magnetic field is in the plane of the sky.  The differences are
unlikely to be detectable with the current generation of X-ray
telescopes.

3.  The \limb\/  model provides a good
description of the limbs of SN~1006, but the \cp\/  and 
\ctr\/ region models overpredict nonthermal emission in other
regions of the remnant if \rolloff\/ is assumed to be constant everywhere.

4. We do find differences in the nonthermal spectrum between the NE
and SW, but they are not sufficient to explain why 
only the NE is detected in TeV $\gamma$-rays, assuming the
$\gamma$-rays are produced by inverse-Compton upscattering of the
cosmic microwave background photons.

The usefulness of \limb, \cp, and \ctr\/ models for other SNRs is not
obvious. Most SNRs have more complex morphology than SN~1006, and few
present regions clearly appropriate for a \ctr\/ or \cp\/ model.
While the \limb\/ model may be appropriate for filamentary structure
at the edge of other SNRs, its curvature differs only very slightly
from the \sresc\/ model. We recall that the model \sresc, and its
derivatives have several assumptions that make them unsuitable for
SNRs of unknown type -- the models presume a Sedov-phase remnant
encountering a uniform upstream medium with a constant magnetic
field. \srcut, as discussed in \citet{Reynolds1999}, which describes
synchrotron radiation with a minimum of assumptions, may be a better
choice for SNRs about whose environment less is known.

\subsection{Thermal Emission}


 1. We can rule out solar abundances in the thermal models in all
regions of the SNR, including regions dominated by nonthermal
emission. 

2. Simple thermal models are incapable of describing the \asca\/ data
adequately.  We note that in much smaller regions of SN~1006, the
analysis of \chandra\/ data encountered similar problems
\citep{Long2003}.  We believe that this is mainly caused by deficient
Fe L-shell atomic data, where many important processes (such as
inner-shell ionization of the Na-like Fe ion) are currently not
included in NEI spectral codes.  It is also likely that the available
spectral models are not adequate for modeling chemically-inhomogeneous
SN ejecta. A full analysis of the combined thermal and nonthermal
emission from SN~1006 will require \xmm\/ and \chandra\/
quality data with substantial improvements in atomic data and
modeling.


3. To the extent that our thermal results can be trusted, we see
elevated abundances which suggest the thermal emission contains
substantial contributions from the reverse shock. Here the models
agree with the silicon equivalent width, and with the results of other
X-ray analyses of SN~1006.












\acknowledgments

We extend special thanks to Rob Roger who, on the eve of his
retirement, rescued the 1988 SN~1006 radio image off a tape, making
radio flux measurements possible.  The Parkes telescope is funded by
the Commonwealth of Australia for operation as a National Facility
managed by CSIRO. The MOST instrument is operated by the University of
Sydney with support from Australian Research Council and the Science
Foundation for Physics within the University of Sidney. This research
has made use of data obtained from the High Energy Astrophysics
Science Archive Research Center (HEASARC), provided by NASA's Goddard
Space Flight Center, NASA's Astrophysics Data System Abstract Service,
and SIMBAD at Centre de Donn\'ees astronomiques de Strasbourg (US
mirror http://simbad.harvard.edu/Simbad). KKD acknowledges support as
a graduate student by NASA grants NAG5-7153 and NGT5-65 through the
Graduate Student Researchers Program (http://education.nasa.gov/gsrp/)
and currently by an NSF Astronomy and Astrophysics Postdoctoral
Fellowship under award AST-0103879. KKD would like to thank
the LHEA staff at NASA's GSFC for their assistance. KJB and SPR thank
the NASA Astrophysics Theory Program grant NAG5-10940 for continuing support.

\clearpage

\begin{figure}
\epsscale{1}
\plotone{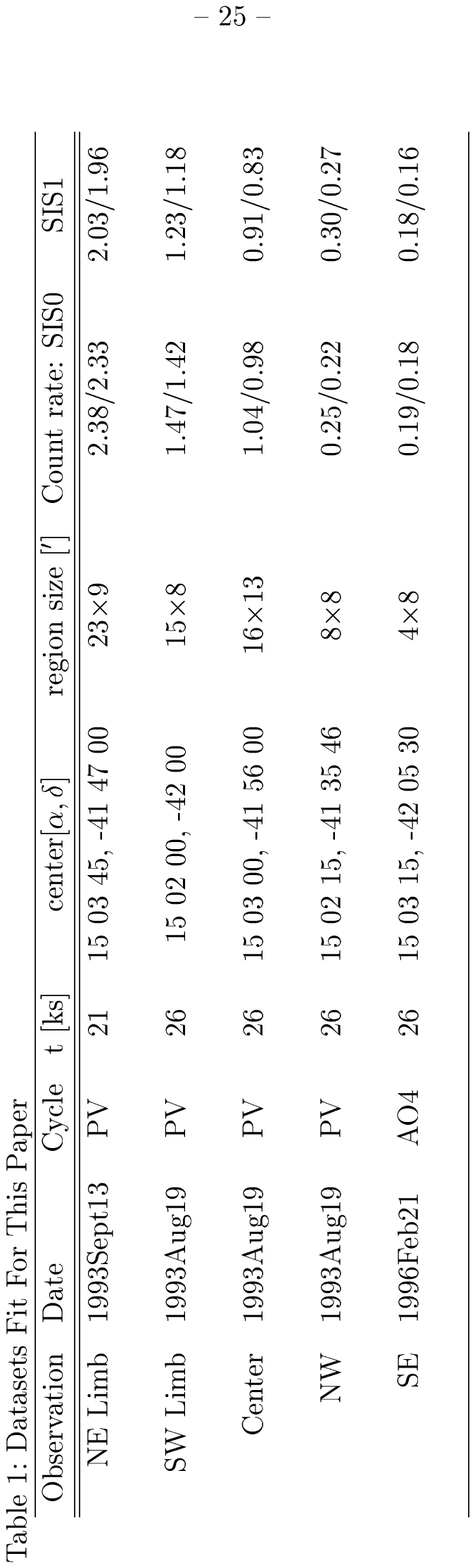}
\caption[]{
\label{tables}}
\end{figure}

\begin{figure}
\epsscale{1.2}
\plotone{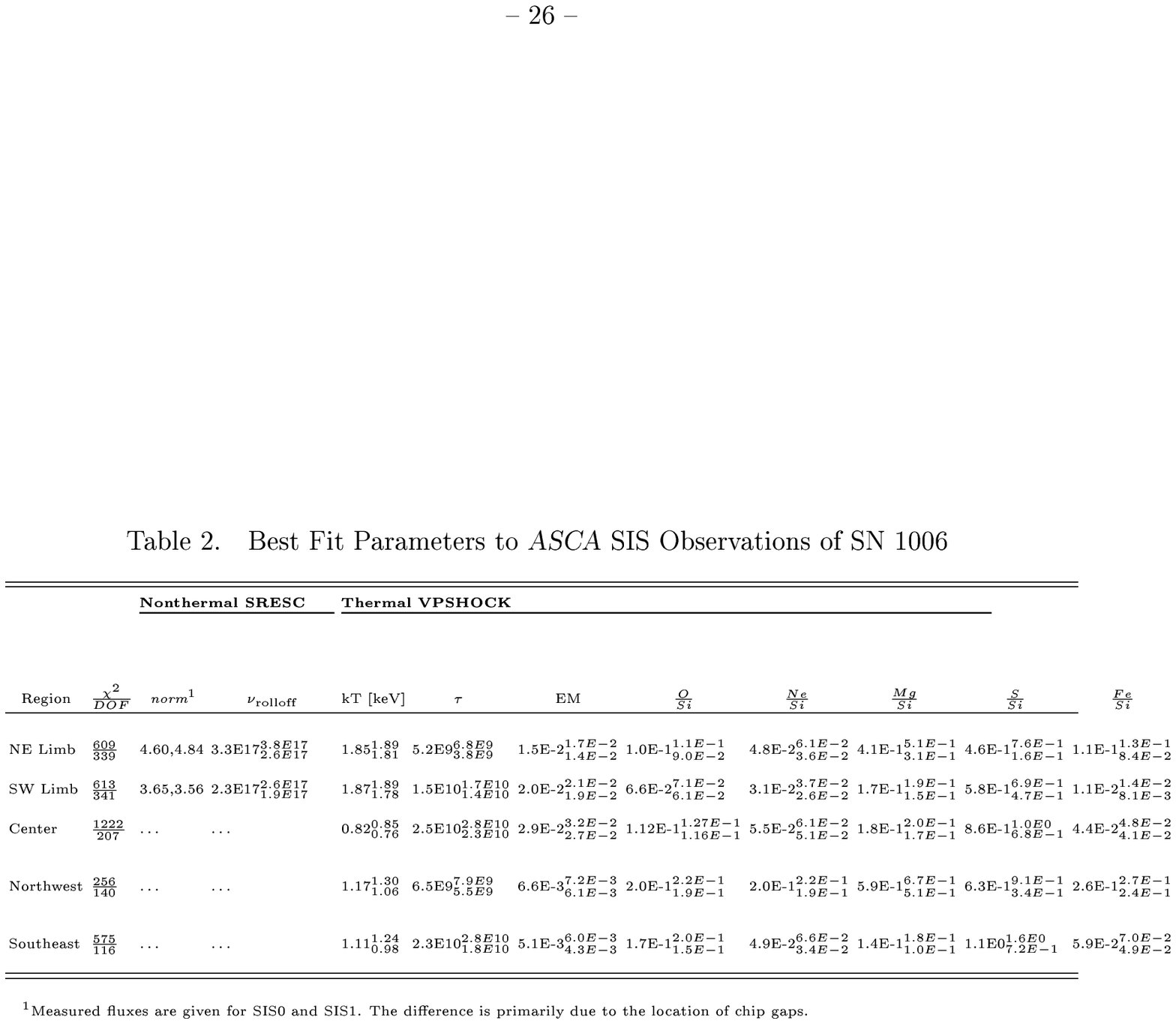}
\caption[]{
\label{tables}}
\end{figure}

\begin{figure}
\epsscale{1}
\plotone{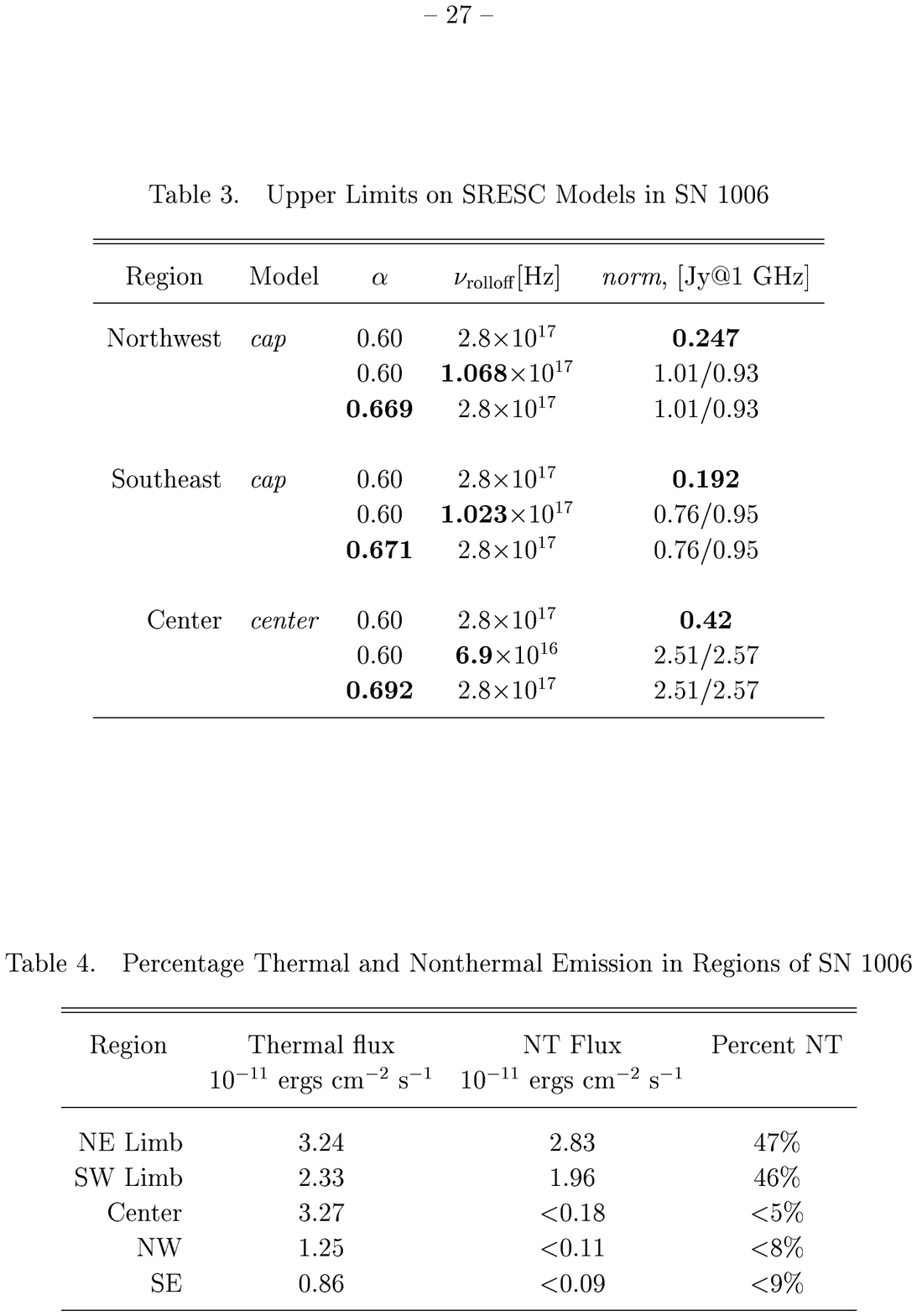}
\caption[]{
\label{tables}}
\end{figure}

\clearpage

\end{document}